\documentclass[]{iopart}
\usepackage{graphicx}
\begin{document}
\bibliographystyle{iopart-num}
\title{Temperature dependence of reflectivity of amorphous silicon dioxide: Evidence of delocalized excitons weakly scattered by phonons}
\author{E. Vella, F. Messina, M. Cannas, R. Boscaino}
\ead{fmessina@fisica.unipa.it}
\address{Dipartimento di Scienze Fisiche ed Astronomiche,
Universit$\grave{a}$ di Palermo, Via Archirafi 36, I-90123 Palermo,
Italy}
\begin{abstract}
We studied the reflectivity spectra of amorphous silicon dioxide detected under vacuum UV synchrotron radiation as a function of temperature between 10 and 300~K. Kramers-Kronig dispersion analysis of reflectivity spectra allowed us to determine the absorption coefficient in the range from 8 to 17.5~eV. Spectra show four main peaks (at $\sim$10.4~eV, at $\sim$11.6~eV, at $\sim$14.1~eV and at $\sim$17.3~eV), the spectral positions of which are consistent with literature data. An appreciable dependence of the line-shape on temperature is observed for the first two peaks only and in particular the peak position of the 10.4~eV resonance red-shifts with increasing temperature above a $\sim$150~K threshold. We demonstrate the exciton peak at 10.4~eV to have a very good Lorentzian band-shape at all the examined temperatures. Based on existing theoretical models, this allows to argue excitons in SiO$_2$ to be weakly scattered by phonons, thus retaining their mobility properties notwithstanding the effects of exciton-phonon coupling and of intrinsic structural disorder of amorphous SiO$_2$. Moreover, the observed temperature dependence of the peak position together with the features of the Urbach absorption tail and of self-trapped exciton emission allow us to estimate the main parameters ruling exciton dynamics in SiO$_2$, such as: the energy of the mean lattice vibrational mode coupled to excitons ($\hbar\omega_0$=0.083~eV), the half width of the excitonic energy band ($B$$\sim$2~eV) and the root mean square amplitude of site-to-site energy fluctuations of exciton energy ($D$$\sim$0.7~eV). The features of the intrinsic Urbach absorption tail can be satisfactorily explained as a consequence of those of the first excitonic peak, supporting the interpretation of the Urbach tail in SiO$_2$ as a consequence of the momentary self-trapping of the 10.4~eV exciton. Finally, the characteristics of the other energy peaks are discussed and an excitonic origin also for the 11.6~eV peak is put forward. On the whole, our results show that exciton dynamics accounts for all optical properties of pure silicon dioxide from $\sim$8 up to $\sim$11~eV.
\end{abstract}
\maketitle
\section{Introduction}
Amorphous silicon dioxide (a-SiO$_2$), or silica, is a solid of primary technologic importance, particularly useful in optical applications requiring a high ultraviolet (UV) transparency. As a consequence, a strong research effort is currently devoted to further improve the optical properties of this material in view of prospective vacuum UV (VUV, $\lambda$$<$200~nm) applications.\cite{Nalwa,SkujaSPIE01,SkujaSPIE03,KajiharaJCSJ07,ChangPRL2000} Moreover, a-SiO$_2$ is an important model system to investigate the physical properties of glassy solids, this study being often aided by the comparison with $\alpha$-quartz (c-SiO$_2$), the most common crystalline polymorph of silicon dioxide. Some of the basic open problems in the physics of glassy SiO$_2$ concern the fundamental absorption edge arising from the electronic transition from the valence to the conduction band. From the experimental point of view, this issue is usually investigated by the combined use of different techniques. Optical absorption spectroscopy allows to characterize the region in which the absorption coefficient $\alpha$ of a bulk solid is smaller than 10$^3$-10$^4$~cm$^{-1}$ and, in systems such as SiO$_2$, it is affected by additional technical complications since the absorption edge falls in the VUV spectral region. In this portion of the spectrum, starting from $\sim$8~eV, both a-SiO$_2$ and c-SiO$_2$ feature the exponential absorption region usually referred to as the Urbach tail.\cite{VellaPRB08} At shorter wavelengths, $\alpha$ becomes exceedingly high to be estimated via standard spectrophotometric measurements in bulk materials, and thus the absorption spectrum can be only indirectly inferred by Kramers-Kronig (K-K) analysis of reflectivity data. Although straightforward in principle, this approach can be quite tricky due to the need of reliable measurements in a sufficiently extended range, obtained by use of a suitable wide band VUV optical source. This has limited the number of K-K investigations on silica and quartz to date,\cite{PhilippSSC66,PhilippJPCS71,LamyAO77,BosioEPL93,TanJACS03,TanPRB05} and consequently hindered the comprehensive understanding of several features, even fundamental, of above-edge absorption in SiO$_2$-based systems.

The exact position and the characteristics of the fundamental electronic transition in SiO$_2$ are still debated albeit a long time-line of experiments devoted to their clarification. Several experimental results led to estimate the band-to-band separation of c-SiO$_2$ and a-SiO$_2$ to be $\sim$9~eV,\cite{WeinbergPRB79,DiStefanoSSC71,PlatzoderPSS68,LaughlinPRB80,ItohPRB89,BosioEPL92,BosioEPL93,TanPRB05} including: the detection of photoconductivity starting from 8.9~eV,\cite{WeinbergPRB79,DiStefanoSSC71} the 8.3~eV photo-excitation threshold of the 2.8~eV luminescence attributed to self-trapped excitons,\cite{ItohPRB89} and the detection in K-K-transformed reflectivity spectra of a region in which absorption is proportional to $(E-E_g)^{3/2}$, with E$_g$=9.3~eV.\cite{BosioEPL92,BosioEPL93} On the other side, some studies questioned these findings, by reporting that photoconductivity and photoemission signals markedly increase only above 11~eV, this being accompanied by qualitative variations of the photoelectron energy distribution curve.\cite{AlexandrovNIMA89,EvrardPRB82,TrukhinJNCS92,Erice} Hence, it was argued that the photoemission signal originally observed below 11~eV was a spurious effect not related to actual band-to-band excitation of the material, while the actual band-gap of SiO$_2$ was proposed to lay around 11~eV.\cite{TrukhinJNCS92} Finally, based on both computational\cite{ChelikowskyPRB77,Pantelides} and experimental results\cite{ItohPRB89} some authors proposed the band-gap at $\sim$9~eV in SiO$_2$ to be indirect, being the threshold for direct band-to-band transitions located at $\sim$11~eV.

The lowest energy peak in the reflectivity spectra of c-SiO$_2$ and a-SiO$_2$ is at 10.3~eV with qualitatively similar characteristics in both systems\cite{LohSSC64,PlatzoderPSS68,Erice,TanPRB05} and it corresponds to a peak in the absorption spectra at $\sim$10.4~eV. This band was promptly attributed to an excitonic transition based on its temperature dependence, as observed in the 300-600~K interval by measurements in c-SiO$_2$.\cite{LohSSC64,PlatzoderPSS68,PhilippJPCS71,SorokinOS76} However, the shape of the 10.4~eV peak was debated for a long time. From the theoretical point of view the broadening of an excitonic resonance in a crystal is a consequence of exciton-phonon interaction. Well established models predict the resulting line shape to be Lorentzian or Gaussian in the limits of weak and strong exciton-phonon scattering respectively,\cite{ToyozawaPTP58,SchreiberJPSJ82} the intensity of which is a function of temperature. The same models are able to predict the Urbach-like character of the far tail of the band at low energies,\cite{VellaPRB08} interpreted as a consequence of momentary exciton self-trapping processes.\cite{Erice,SumiJPSJ71,SchreiberJPSJ82c} Although these models were not originally proposed for disordered systems, it was pointed out that the effects of a small structural disorder in a crystalline system are indistinguishable from those of thermal fluctuations.\cite{ToyozawaPTP58} From the experimental point of view, some of the existing data suggested the situation in SiO$_2$ to be far more intricate than above described. Indeed, photoluminescence measurements upon above-gap excitation of c-SiO$_2$ and a-SiO$_2$ doped with luminescent impurities have led some authors to distinguish different types of excitons based on their different mobility and energy transfer properties.\cite{AlexandrovNIMA89,TrukhinJNCS92,Erice} Being also the width of the $\sim$10.4~eV absorption peak in SiO$_2$ significantly wider than in other crystals, it was proposed that such band actually arises from the superposition of several unresolved Gaussian components, corresponding to different exciton types.\cite{AlexandrovNIMA89,TrukhinJNCS92,Erice}

Besides the 10.4~eV peak, the origin and properties of the other features evidenced by reflectivity spectra are even more debated. Based on the temperature dependence, the 11.5~eV peak was argued to contain two sub-structures, one of which supposedly excitonic.\cite{PlatzoderPSS68,BosioEPL92} However, other authors have endorsed the interpretation of the 11.5~eV peak as a maximum of the inter-band transition cross section.\cite{SorokinOS76,TanPRB05} Finally, the higher energy features (14.0~eV and 17.5~eV) are almost universally regarded as inter-band transition peaks,\cite{TanPRB05} although contrasting experimental evidences were reported about their temperature dependence: they were found either to be independent of temperature\cite{PlatzoderPSS68} or to shift to higher energies by 0.1~eV when going from 300 to 30~K.\cite{BosioEPL92}

In this work we contribute to the clarification of these topics by providing a comprehensive investigation of excitonic absorption and emission in SiO$_2$ founded on reflectivity and photoluminescence measurements. We have recently proposed that Toyozawa's model can be successfully applied to a-SiO$_2$ as well, where the 10.4~eV peak turns out to be Lorentzian leading to the conclusion that excitons retain their delocalized properties albeit thermal and structural disorder.\cite{MessinaPRL2010} Now, based on a detailed analysis of the optical properties of SiO$_2$ in the VUV range, and on a critical comparison with existing literature data, the present work deepens this interpretation of exciton physics in SiO$_2$ and introduces a self-consistent description encompassing the features of the excitonic peaks and those of the low energy Urbach absorption tail. Furthermore, we place our results in the wider context of excitonic properties in wide band-gap solids, showing that, as in systems like LiF, excitons in SiO$_2$ are weakly scattered by, but strongly coupled to phonons.
\section{Theoretical description of excitonic optical properties}\label{theory}
In absence of exciton-phonon interaction an excitonic absorption line in a crystal would be infinitely sharp and located at the energy $E_P^0$ corresponding to the creation of the exciton with wave number $K$$=$$0$, in order to fulfill the conservation of crystalline momentum in photon absorption. The well established theory by Toyozawa and coworkers describes under certain basic assumptions the modifications of exciton optical properties arising from the quasi-particle interaction with the phonon bath.\cite{ToyozawaPTP58,Toyozawa} It turns out that when exciton-phonon coupling (EPC) is taken into account, the resulting absorption line-shape is basically governed by the competition between the mobile nature of excitons and their tendency to localization in the lattice due to EPC. The former is measured by the half-width $B$ of the energy band of excitons, proportional to site-to-site transfer rate, while the latter is measured by the root mean square amplitude $D$ of the fluctuations of exciton energy due to the thermal activated vibrations of the lattice. Theory predicts that if the ratio $D/B$$\ll$1 and the mobile nature of the exciton is thus prevailing (weak scattering limit), the exciton absorption line-shape is Lorentzian:\cite{ToyozawaPTP58,Toyozawa,SchreiberJPSJ82}
\begin{equation}
\alpha(E,T)=\alpha_0\left[1+\left(\frac{E-(E_P^0+\Delta_0(T))}{\Gamma_0(T)}\right)^2\right]^{-1}
\label{LorentzianProfile}
\end{equation}
While the delocalized nature of exciton states is not substantially altered by weak phonon scattering, the Lorentzian broadening is ultimately due to the fact that the $K$$=$$0$ exciton acquires a finite lifetime due to phonon-induced scattering towards $K$$\neq$$0$ levels. This situation can occur either if EPC is inefficient (small $D$) or if excitons are mobile enough (large $B$) to compensate for its effects. In this case, the peak position $E_P^0$ is perturbed by the self-energy $\Delta_0(T) +i\Gamma_0(T)$ of the exciton in the phonon field, giving rise to a temperature dependent shift ($\Delta_0(T)$) and broadening ($\Gamma_0(T)$) of the exciton line. The predicted FWHM=2$\Gamma_0(T)$ can be expressed as $cD^2/B$, where the adimensional proportionality coefficient $c$ is the parameter most affected by the details involved in theoretical modeling of exciton-phonon interaction and by the detailed features of the exciton band. If the $K$$=$$0$ level is neither at the bottom nor at the top of the excitonic band, the first-order value of $c$ turns out to be $\sim\pi$.\cite{ToyozawaPAC1997} The quantities $D$, $\Delta_0(T)$ and $\Gamma_0(T)$ are expected to increase with growing temperature due to progressive thermal activation of the lattice. Indicating with $\hbar\omega_0$ the energy of the mean phonon mode coupled to the exciton, the predicted temperature dependencies are in the form:
\begin{equation}
\frac{\Delta_0(T)}{\Delta_0^0}=\frac{\Gamma_0(T)}{\Gamma_0^0}=\frac{D^2}{\hbar\omega_0 E_{LR}}=\coth\left(\frac{\hbar\omega_0}{2k_BT}\right)\label{allvsT}
\end{equation}
where the parameter $E_{LR}$ can be interpreted as the lattice relaxation energy after the creation of an exciton due to vibrational decay within the excited electronic state.

As $D$ grows with temperature, the effect of EPC on the excitonic absorption line increases, until it cannot be considered anymore as a small perturbation affecting the delocalized exciton states. In particular, for thermal or structural disorder so large, or for $B$ so small, that $D/B$$\gg$1 (strong scattering limit), the model foresees the overall exciton line-shape to be Gaussian with $\sim$2.35$D$ FWHM.\cite{ToyozawaPTP58,Toyozawa,SchreiberJPSJ82} This case corresponds to a completely localized, defect-like exciton, whose absorption line basically reproduces the density of states whose energy thermally fluctuates with Gaussian statistics from site-to-site.

It is worth noting that the transition between delocalized ($D$$\ll$$B$) and localized ($D$$\gg$$B$) exciton must be regarded as a gradual one as a function of increasing thermal disorder. On the other side, the relaxation pathways of the exciton returning to the ground state after excitation are predicted to feature an almost abrupt transition between two situations: excitons which go back to the ground state by emitting a resonant sharp emission line still associated to a free exciton (FE) and systems emitting the typical broad Stokes-shifted band due to self-trapped excitons (STE). The transition is controlled by the value of the so-called EPC constant $g$$=$$E_{LR}/B$: if $g$$>$$1$ (strong EPC) exciton emission is of the self-trapped type at energy $\sim E_P^0-2E_{LR}$, while $g$$<$$2/3$ (low EPC) leads to free-exciton emission without significant Stokes shift from the absorption line.\cite{SumiJPSJ71,SchreiberJPSJ82c,ToyozawaPAC1997,Toyozawa} The interval $2/3$$<$$g$$<$$1$ allows for the coexistence of STE and FE emissions. The strong EPC condition is associated to the existence of an energetically stable self-trapped exciton state to which the exciton can relax upon a lattice rearrangement. Since the boundary between weak and strong scattering ($D$/$B$=1) is different from that between weak and strong coupling ($g$=1), strong EPC does not necessarily imply strong scattering and viceversa: in particular, values of $B$, $E_{LR}$, $D$ are possible for which strong coupling and weak scattering coexist. In this situation, the exciton behaves as a nearly free quasi-particle until it finally surmounts the potential barrier to become self-trapped.\cite{ToyozawaPAC1997}

Finally, Toyozawa's model is even able to successfully predict the existence of the exponential absorption tail, experimentally observed in a wide variety of materials, known as Urbach rule,\cite{Tauc,Toyozawa} and usually treated only from a phenomenological point of view. Actually in the frame of Toyozawa's theory such a tail naturally arises in the far low energy wing of a weak scattered Lorentzian excitonic absorption peak. In fact, even in the weak scattering case, EPC induces the appearance of localized states below the bottom of the excitonic band which can be regarded as temporary self-trapped exciton states showing up from time to time due to sufficiently great local lattice deformations. The existence of localized states below delocalized ones in the thermally-disordered lattice can be actually considered as a form of Anderson localization.\cite{Toyozawa} Thus, while the main Lorentzian absorption band is associated to the delocalized states, excitation of the localized states of lower energy gives rise to an absorption wing which has been numerically predicted to be exponential:\cite{SchreiberJPSJ82c,Toyozawa}
\begin{equation}
\alpha(E)=\alpha_0\exp{\left(-\sigma\frac{E_0-E}{k_B T}\right)}\label{urbach}
\end{equation}
where $\sigma$ is expected to depend on temperature according to:
\begin{equation}
\sigma=\sigma_0 \frac{2k_BT}{\hbar\omega} \tanh{\left(\frac{\hbar\omega_0}{2k_BT}\right)}\label{sigma}
\end{equation}
The parameter $\sigma_0$ controlling the slope of the exponential tail is related to the EPC constant: $\sigma_0=s/g$ where $s$ is a dimensionless parameter of the order of unity depending on geometrical factors such as lattice structure.\cite{SchreiberJPSJ82c,Toyozawa} This relation allows to estimate the EPC constant $g$ from absorption measurements in the region of the Urbach tail and thus to predict the existence of STE (if $g$$>$1) or FE (if $g$$<$1) emission. This prediction has been verified to be correct for a great number of materials.\cite{Toyozawa}
\section{Materials and methods}
\subsection{Experimental techniques}
We performed reflectivity measurements both on a-SiO$_2$ and c-SiO$_2$ samples. The a-SiO$_2$ samples were commercial type IV synthetic samples (commercial nickname: Suprasil F300, trademark of Heraeus Quarzglas,\footnote{http://heraeus-quarzglas.com/en/home/Home.aspx} OH content$<$1 part per million in weight), cylindrically shaped with 5~mm diameter and 2~mm thickness, and produced by reaction of O$_2$ with SiCl$_4$ in water-free plasma. The c-SiO$_2$ sample was a z-cut $\alpha$-quartz sample 5$\times$5$\times$1~mm$^3$ in sizes. All samples were subjected to industrial grade optical polishing on major surfaces and to a three-steps ultrasonic cleaning (10' in acetone, 10' in ethanol and 10' in deionized water) before measurements. Measurements were carried out upon excitation with synchrotron radiation in the photon energy range from 8.0 to 22.5~eV at the SUPERLUMI experimental station in DESY - Hamburg (HASYLAB - Beamline I). The excitation wavelength was selected via a 2-m monochromator. The accuracy of the measured wavelength was 0.1~nm in the whole explored range. We verified that stray light and second-order contributions to the exciting beam can be safely neglected throughout the whole chosen spectral range. The excitation beam had a 2.5~mm$^2$ cross section and hit the samples with a 17.5$^{\circ}$ angle of incidence on one of their major surfaces. The reflected beam was detected by a photomultiplier coated by sodium salicylate (C$_7$H$_5$NaO$_3$) on the entry window, in order to ensure flat spectral response. All the experimental system was kept in high vacuum (10$^{-9}$-10$^{-8}$~mbar). Each acquired spectrum was scaled for the spectral density curve of the excitation beam, measured prior to the experiments by recording on a second photomultiplier the photoluminescence excitation spectrum of a sodium salicylate reference sample. Temperature was controlled by a liquid He flow cryostat.

Photoluminescence emission spectra under excitation with synchrotron radiation were acquired as well. The emission spectra were detected in front-face geometry. The signal was dispersed by a 300~lines/mm monochromator with 500~nm
blaze, and detected by a charge coupled camera cooled by liquid N$_2$. The emission bandwidth was 10~nm. Spectra were corrected for the response and dispersion of the detection system.
\subsection{Kramers-Kronig analysis}
We describe now the Kramers-Kronig analysis performed on reflectance data in order to extract the absorption spectra. Reflection of a monochromatic beam with photon energy $E$ by a dielectric is governed by the complex reflection coefficient $r$, which is related to the complex refractive index $n+ik$ by:\cite{Ashcroft}
\begin{equation}
r=\sqrt{R(E)}e^{i\theta(E)}=\frac{n-1+ik}{n+1+ik} \label{rvsn}
\end{equation}
where $R(E)$, the square modulus of $r(E)$, physically corresponds to the measured reflectance. In using Eq.~\ref{rvsn}, we are implicity approximating the angle of incidence to zero. One of the forms in which K-K relations may be written allows to express the phase $\theta(E)$ in terms of the modulus $\sqrt{R(E)}$:
\begin{equation}
\theta(E)=-\frac{2E}{\pi}P\int^{\infty}_0\frac{ln\sqrt{R(E')}}{E'^2-E^2}dE' \label{KK}
\end{equation}
Where $P$ indicates the Cauchy principal value of the integral. Hence, starting from the experimental $R(E)$, one can use Eq.~\ref{KK} to calculate $\theta(E)$ and eventually $n$ and $k$ by inverting Eq.~\ref{rvsn}. The imaginary part $k$ of the refractive index is proportional to the absorption coefficient $\alpha(E)$:
\begin{equation}
\alpha(E)=\frac{2Ek(E)}{\hbar c} \label{alphak}
\end{equation}
In order to obtain reliable results from numerical K-K integration, the $R(E)$ function used as an input to the integral in Eq.~\ref{KK} must be known from $E$=0 to a sufficiently high energy to allow the integral to converge. Since in practice any experimental estimation of $R(E)$ is available in a limited interval $E_m$$<$$E$$<$$E_M$, prior to integration it is mandatory to carry out an extension of experimental data on the low and high energy tails.

In regard to the low energy side, we used literature data reported by Philipp to extend data down to 0~eV.\cite{PhilippJPCS71} To this aim, we scaled each of our spectra for a constant properly chosen to lock the measured value of $R(E_m)$ to Philipp's, and we used Philipp's data for $E$$<$$E_m$. As a side effect, this procedure automatically converts data from relative to absolute reflectance measurements. It turns out that our reflectivity spectra are in a very good agreement with Philipp's data in the whole 9-10~eV region, so that $E_m$ can be chosen rather arbitrarily within this interval without affecting the results. Data reported later on in the paper were calculated with $E_m$=9.3~eV. We used Philipp's data at room temperature for all spectra, based on the fact that the refractive index of SiO$_2$ in the visible and UV regions undergoes only very small ($\sim$0.003\%) variations between 10 and 300~K.\cite{LevitonSPIE2006,MatsuokaJNCS1991}

On the high energy side the extension was done analytically, as customary, by making use of a closure relation based on theoretical grounds, the standard choice being the free-electron gas approximation $R(E)\propto E^{-4}$. Hence we used a $aE^{-4}$ equation for $E$$>$$E_M$, the value of the $a$ constant being chosen for each spectrum so as to ensure continuity with the experimental value $R(E_M)$. The specific value of $E_M$=21~eV was chosen as it led to the best agreement between the first derivative of the spectrum and that of the $E^{-4}$ curve. To estimate the size of the errors introduced by K-K analysis we verified the influence of different choices for the high-energy extrapolation on the line-shape of the 10.4~eV peak in the absorption spectra. We considered, in particular, the influence of the changes on both the exponent of the free-electron gas approximation curve, $E^{-b}$, and the energy threshold, $E_M$, above which the $E^{-b}$ curve was used. We repeated K-K integration with $b$=3.8 and $b$=4.2. Only very small differences were observed and we verified that in each of the three cases the peak kept the same line-shape, although slight variations in the peak position (within 20~meV) and width (within 7$\%$) were seen.
As far as the choice of the energy threshold is concerned, we verified that repeating the calculation for $E_M$=19~eV and 20~eV the changes on the 10.4~eV peak were almost not appreciable. In fact, although some variations in the resulting absorption spectra were obviously observed in the region close to the $E_M$ value, in the range up to 17.5~eV the spectra obtained for the three different $E_M$ values were closely overlapping. For this reason in the following we will report the absorption coefficient up to 17.5~eV.

After preparation of the $R(E)$ curves, we numerically performed K-K integration up to 3000~eV, and finally calculated $\alpha(E)$ via Eqs.~\ref{rvsn} and \ref{alphak}. We verified that no appreciable variations of the obtained $n$ and $k$ occur if one further extends the upper numerical integration limit above 3000~eV.
\section{Results}
\begin{figure}
\includegraphics[width=8cm]{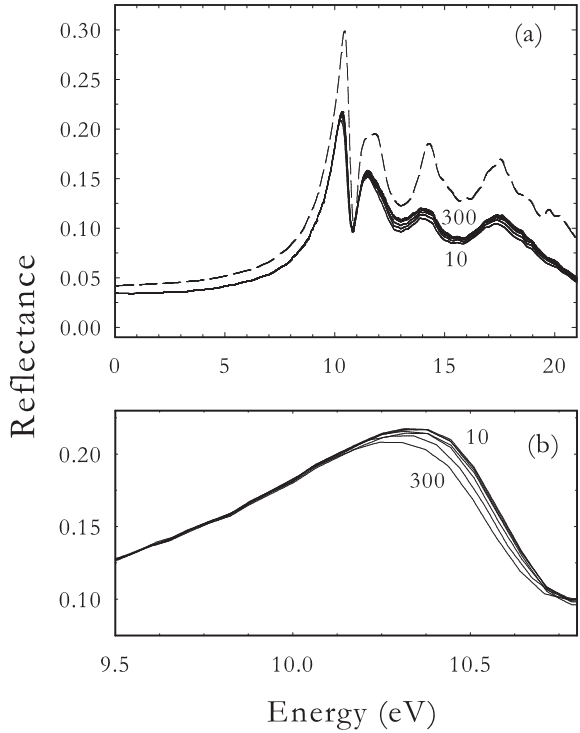}
\caption{\label{SpettriRifl}Panel (a): reflectivity spectra of a-SiO$_2$ (full lines) in the temperature range from 10 to 300~K by steps of 50~K and of c-SiO$_2$ (dashed line) at 10~K. Panel (b): Enlargement of the peak at $\sim$10.4~eV of a-SiO$_2$. Labels represent temperatures (K) of the closest curve.}
\end{figure}
Figure~\ref{SpettriRifl}-(a) shows the reflectivity spectra of a-SiO$_2$ in the range from 0 to $\sim$21~eV. Spectra were acquired in the range of temperatures from 10 to 300~K by steps of 50~K. The spectrum of c-SiO$_2$ at 10~K is reported for comparison. Spectra at all temperatures are characterized by the presence of four main peaks: a first relatively narrow and more intense peak at $\sim$10.3~eV and three other broader peaks at $\sim$11.5, $\sim$14.0 and $\sim$17.3~eV respectively. The positions of the peaks in c-SiO$_2$ are $\sim$10.4~eV, $\sim$11.4~eV, $\sim$14.1~eV, $\sim$17.3~eV.
The reflectivity spectra are significantly affected by temperature. In particular, detailed inspection of the spectra reveals that whereas the peaks at $\sim$10.3 and $\sim$11.5~eV change in both their positions and amplitudes as a function of temperature, the peaks at $\sim$14.0 and $\sim$17.3~eV increase their intensities without noticeably changing their shapes with increasing temperature, with all the changes occurring from 10 to 150~K only. In Fig.~\ref{SpettriRifl}-(b) an enlargement of the peak at $\sim$10.3~eV of a-SiO$_2$ in the region from 9.5 to $\sim$10.7~eV is reported: as it is possible to see, with increasing temperature the peak red-shifts while its peak amplitude decreases.
\begin{figure}
\includegraphics[width=8cm]{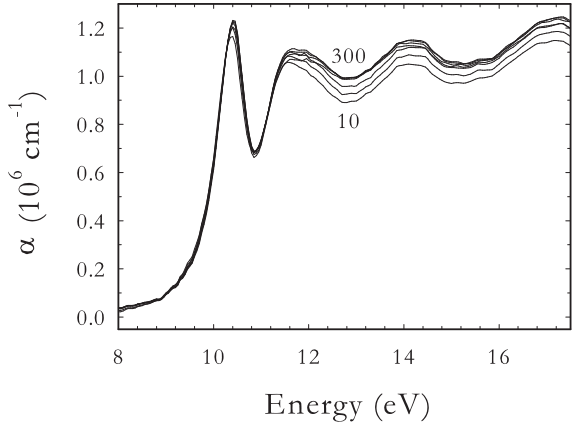}
\caption{\label{SpettriAlfa}Absorption coefficient of a-SiO$_2$, as estimated via K-K analysis in the temperature range from 10 to 300~K by steps of 50~K. Labels represent temperatures (K) of the closest curve.}
\end{figure}
\begin{figure}
\includegraphics[width=8cm]{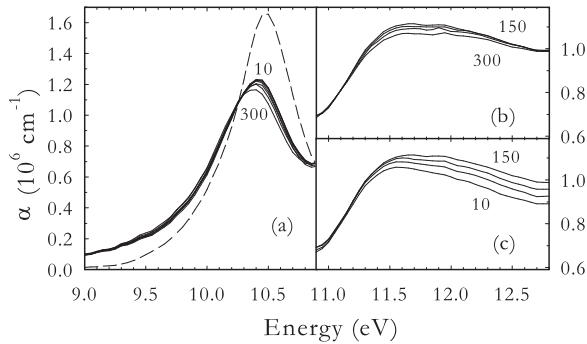}
\caption{\label{SpettriAlfaZoom}(a) Enlargement of the absorption peak at $\sim$10.4~eV from 10 to 300~K by steps of 50~K (full lines). Absorption coefficient of c-SiO$_2$ at 10~K in the same spectral region (dashed line) (b) Enlargement of the peak at $\sim$11.6~eV in a-SiO$_2$ from 300 to 150~K by steps of 50~K. (c) Enlargement of the peak at $\sim$11.6~eV in a-SiO$_2$ from 150 to 10~K by steps of 50~K. Labels represent temperatures (K) of the closest curve.}
\end{figure}

In Fig.~\ref{SpettriAlfa} the absorption coefficient of a-SiO$_2$ is presented in the range from 8 to 17.5~eV. The absorption spectra were obtained from the reflectivity spectra shown in Fig.~\ref{SpettriRifl} using the Kramers-Kronig analysis described in the previous section. As it can be seen, four main peaks are present as well: at $\sim$10.4~eV, at $\sim$11.6~eV, at $\sim$14.1~eV and at $\sim$17.3~eV. In order to evidence the dependence of the features of the peaks on temperature, in Fig.~\ref{SpettriAlfaZoom} enlargements of both peaks at $\sim$10.4~eV and at $\sim$11.6~eV are shown. In Fig.~\ref{SpettriAlfaZoom}-(a) the absorption spectra in the range from 9 to 11~eV are presented: with increasing temperature the peak at $\sim$10.4~eV shifts towards lower energies. The enlargement in the region of the $\sim$10.5~eV peak of the absorption spectrum of c-SiO$_2$ at 10~K obtained from Kramers-Kronig analysis is reported for comparison. The changes of the peak at $\sim$11.6~eV are not monotonous with temperature, as evident from the enlargements of the spectra in the range from $\sim$11 to $\sim$13~eV shown in Fig.~\ref{SpettriAlfaZoom}-(b) and Fig.~\ref{SpettriAlfaZoom}-(c), where the absorption spectra from 300 to 150~K and from 150 to 10~K are reported respectively. In the range from 300 to 150~K (panel (b)), absorption at $\sim$11.6~eV decreases with increasing temperature whereas in the range from 150 to 10~K (panel (c)) it grows with increasing temperature. As far as the two remaining peaks at $\sim$14.1 and $\sim$17.3~eV are concerned, spectra in Fig.~\ref{SpettriAlfa} show that in the range from 10 to 150~K their overall intensities increase, whereas for higher temperatures they do not change noticeably.
\begin{figure}
\includegraphics[width=8cm]{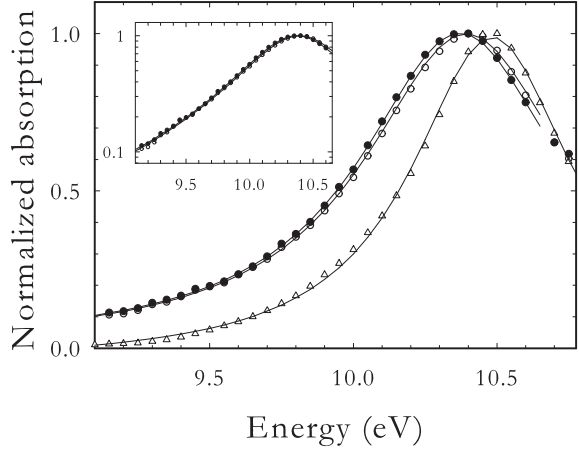}
\caption{\label{PrimoPiccoFitLoren} Normalized absorption at 10~K (open circles) and at 300~K in a-SiO$_2$ (filled circles), and at 10~K in c-SiO$_2$ (triangles). Continuous lines were obtained via a Lorentzian fit of each spectrum. Inset: semilogarithmic plot of the same data on a-SiO$_2$ as in the main panel.}
\end{figure}

Figure~\ref{PrimoPiccoFitLoren} presents the normalized absorption spectra of a-SiO$_2$ at T=10~K and 300~K respectively, and that of c-SiO$_2$ at 10~K, in the energy range from 9.1 to 10.8~eV. As evident, the peak in a-SiO$_2$ red-shifts with increasing temperature without appreciably modifying its shape. The peak at $\sim$10.4~eV in SiO$_2$ turns out to have a very good Lorentzian profile at all temperatures. The Lorentzian best fit curves (solid lines) of a-SiO$_2$ spectra at T=10~K and 300~K are reported in Fig.~\ref{PrimoPiccoFitLoren} as well: as it is possible to see, their agreement with experimental data is very close even on the left band wings. Fittings were performed on the spectra in the range from 9 to 10.6~eV in a-SiO$_2$ and from 9 to 10.8~eV in c-SiO$_2$. In order to evidence the good quality of the fitting procedure of the experimental spectra, in the inset of Fig.~\ref{PrimoPiccoFitLoren} the same spectra and best fit curves are shown in a semi-logarithmic scale. Finally, the figure shows that also the peak in c-SiO$_2$ can be fitted well by a Lorentzian curve.
\begin{figure}
\includegraphics[width=8cm]{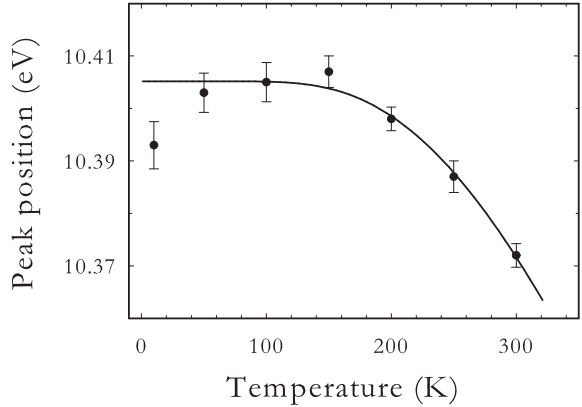}
\caption{\label{PiccoEIntens_vs_T} Position of the peak at $\sim$10.4~eV in a-SiO$_2$ as a function of temperature, as determined via a Lorentzian fit of the peak profiles. The full line is discussed in the text.}
\end{figure}

Figure~\ref{PiccoEIntens_vs_T} shows the position of the peak at $\sim$10.4~eV as a function of temperature, as obtained by the Lorentzian fitting. The position of the peak remains approximately constant around a value of $\sim$10.41~eV from 10 to 150~K, whereas it shifts to lower energies for higher temperatures. No change of the full width at half maximum (FWHM) of the peak is appreciable within experimental error: its value is $(0.85\pm0.08)$~eV. In c-SiO$_2$ at 10~K the FWHM of the 10.5~eV peak is $(0.69\pm0.04)$~eV. In the matter of the peak at $\sim$11.6~eV, its position and amplitude, or area, cannot be quantitatively determined due to its closeness to the two other peaks at $\sim$10.4 and $\sim$14.1~eV respectively.

As a final remark, we point out that present data cannot be used to directly evaluate the parameters of the Urbach tail typical of the $E$$<$9~eV spectral region. In fact, the noise amplitude on the spectra in Fig.~\ref{SpettriAlfa} is of the order of 10$^3$~cm$^{-1}$, and a constant contribution to absorption of the same order of magnitude is also possible due to effects such as incomplete subtraction of dark current from raw reflectance data. Hence, K-K absorption spectra are reliable only for absorption coefficients larger than about 10$^4$~cm$^{-1}$. While this is only 1\% of the 10.4~eV peak amplitude and it does not affect our analysis for E$>$9~eV, on the other hand it means that the Urbach tail is concealed within experimental error on the K-K absorption spectrum, as usual for this kind of analysis. It is impossible, then, to analyze here the spectral region in which the Urbach tail joins the left wing of the excitonic peak, where $\alpha(E)$ is expected to feature an inflection point. For this reason, in the discussion section the properties of the 10.4~eV peak will be compared with those of the Urbach tail as arising from previous studies.
\section{Discussion}
This section is devoted to the discussion of our experimental results and is organized as follows. In subsection~\ref{sectionA} we discuss the mobility properties of excitons in SiO$_2$ summarizing what we have already reported in a recent paper.\cite{MessinaPRL2010} In subsection~\ref{urbachsection} we analyze the relations existing between the features of the first excitonic peak and those of the Urbach absorption tail. Subsection~\ref{parameters} is devoted to providing estimates of the physical parameters ruling exciton dynamics in SiO$_2$. Subsection~\ref{sectionD} deals with the comparison between our results and literature data on SiO$_2$ and on other solid systems. Finally in subsection~\ref{sectionE} the properties of the peaks at energies higher than 10.4~eV are discussed.

\subsection{Comparison with theoretical models: Exciton mobility in SiO$_2$}\label{sectionA}
The room temperature reflectivity spectrum, shown among the others in Fig.~\ref{SpettriRifl}, agrees qualitatively with those observed in other a-SiO$_2$ samples and previously reported in literature.\cite{PhilippJPCS71,BosioCM1989,BosioEPL92,TanPRB05} The positions of the reflectivity peaks, $\sim$10.3, $\sim$11.5, $\sim$14.0 and $\sim$17.3~eV, are roughly the same as reported before, as it can be said for the absorption peaks estimated by K-K dispersion analysis. It is widely accepted in literature that in both c-SiO$_2$ and a-SiO$_2$ the 10.4 eV peak has an excitonic origin.\cite{LohSSC64,PlatzoderPSS68,SorokinOS76,AlexandrovNIMA89,TrukhinJNCS92,BosioEPL93,TanPRB05} We have shown in Fig.~\ref{PrimoPiccoFitLoren} that this peak has a remarkably good Lorentzian line-shape at all the examined temperatures, as already reported in a recent paper.\cite{MessinaPRL2010} This circumstance bears a deep meaning within Toyozawa's theory of line-shapes of excitonic absorption bands, in the limit of weak exciton-phonon scattering.\cite{ToyozawaPTP58,SchreiberJPSJ82} First our analysis shows that this theory can be extended to an amorphous system, and this allows us to argue that the effect of structural disorder intrinsic to the amorphous a-SiO$_2$ structure is qualitatively the same as that of increasing temperature, in that both lead to an increase of the root mean square amplitude $D$ of excitonic energy fluctuations (see section~\ref{theory}). As a matter of fact this point of view had already been suggested to apply to the effect of structural disorder introduced in a crystal by dislocations, vacancies, interstitial atoms or impurities.\cite{ToyozawaPTP58} Second and most important, the close Lorentzian line-shape proves the mobile, delocalized nature of excitons in both c-SiO$_2$ and a-SiO$_2$. Present data clearly point out that from 10 to 300~K excitons mobility is so large as compared to the effects of both static disorder and EPC in a-SiO$_2$ that the exciton band keeps its Lorentzian profile, not presenting any tendencies towards a Gaussian one even at the highest considered temperature (300~K). This result is particularly remarkable in a-SiO$_2$ because exciton mobility is ultimately related to the translational symmetry of the lattice. Thus, the lack of long range order typical of an amorphous solid turns out to have a negligible effect on the properties of excitons, meaning that the effect of static disorder in a-SiO$_2$ is low enough to closely preserve crystal-like delocalization of excitons. It appears that, at least at low temperature, the difference between the excitonic properties of a-SiO$_2$ and c-SiO$_2$ comes down to a greater width of the first exciton absorption peak in the amorphous system, $(0.85\pm0.08)$~eV instead than $(0.69\pm0.04)$~eV, likely due to structural disorder introducing small additional site-to-site fluctuations of the exciton energy contributing to $D$.

It is worth noting that besides Toyozawa's model, very general theoretical arguments strongly suggest that Lorentzian absorption line-shapes are expected when excitonic mobility effects ($B$) prevail over site-to-site energy fluctuations ($D$). These arise from the general theoretical description by Kubo of the broadening of a resonance due to a random fluctuation of its proper energy.\cite{KuboJMathPhys63} Assuming the resonance energy to undergo Gaussian fluctuations of amplitude $D$, the broadening can be shown being either Lorentzian or Gaussian depending on the relation between $D$ and the correlation time $\tau$ of the fluctuations: if $D$$\ll$$\hbar/\tau$ the line-shape is Lorentzian, whereas in the opposite case it is Gaussian. From a physical point of view the statistical fluctuations of lattice positions are a straightforward source of fluctuations (having amplitude $D$) of the excitonic energy on a given site, while the rate $B$ of site-to-site hops of excitons entails a correlation time $\hbar/B$ of the fluctuations experienced by excitons (if one neglects the inter-site correlations). Within this scheme, the condition $D$$\ll$$\hbar/\tau$ boils down to $D/B$$\ll$1. These arguments are particularly strong and useful since they rely on more general bases and thus they do not depend on the detailed theoretical modeling of the exciton-phonon scattering process.

The close agreement between experimental data and a simple 3-parameters Lorentzian function without the help of any baseline virtually proves that any other contribution to absorption aside from that due to the 10.4~eV exciton must be negligible in the 9-10.6~eV fitting range. On the other side, from the physical point of view one may expect fundamental non-excitonic band-to-band absorption to introduce a non-constant baseline possibly influencing our analysis. In this context the most reliable way to assess the influence of fundamental absorption is to infer its shape from actual experimental photoconductivity data available in literature.\cite{Erice} The energy dependence of photoconductivity detected in c-SiO$_2$ turns out to be sigmoid-shaped being very low below about 11~eV and joining the absorption profile at about 15~eV. Thus, existing experimental data consistently confirm that the influence of band-to-band absorption is negligible on the 10.4~eV absorption peak line-shape.\footnote{Aiming to be very conservative, we evaluated anyway the possible effect of a baseline represented by a cubic function and joining the absorption spectrum at the right side of the peak near 11~eV. By subtracting such a baseline to the spectrum of a-SiO$_2$ at 10~K, and repeating a Lorentzian fit on the residual peak, it turns out that the latter is still consistent with a Lorentzian profile within experimental error, while it is definitely not Gaussian. Lorentzian fitting yields a 0.74~eV width, as opposed to 0.85~eV. Thus, even in this case, the only effect of a baseline subtraction would  possibly be a reduction of the band-width, not affecting significantly the Lorentzian line-shape.}

Let us now discuss the temperature dependence of the spectral features of the 10.4~eV peak. The position of the peak as a function of temperature is expected to be described by the equation: $E_P(T)=E_P^0+\Delta_0(T)$, $\Delta_0(T)$ being given by Eq.~\ref{allvsT}. In Fig.~\ref{PiccoEIntens_vs_T} the values of the peak positions as determined from the Lorentzian fittings of experimental spectra and the curve obtained by least-square fitting data with this equation are reported. As it can be seen, Eq.~\ref{allvsT} is consistent with the observed trend, with the only possible exception of the point at T=10~K. Admittedly present results do not allow to establish if the observed scatter of this datum points out a low temperature deviation from the accepted theoretical description as far as the temperature dependence of the peak is concerned. The best fitting values of the parameters are: $E_P^0$=$(10.79\pm0.05)$~eV, $\Delta_0^0$=$(-0.38\pm0.05)$~eV and $\hbar\omega_0$=$(0.083\pm0.002)$~eV and are reported in Table~\ref{table:res}. We expect the width of the peak to have the same temperature dependence as that of the peak position, i.e. governed by Eq.~\ref{allvsT} as well. Under this assumption and using for $\hbar\omega_0$ the value reported above, the expected variation of the width of the band in the range of temperatures from 10 to 300~K is 8$\%$, corresponding to an absolute variation of 0.07~eV. Unfortunately, this is smaller than our 0.08~eV experimental error on the FWHM, thus preventing us to directly detect the progressive broadening in the investigated temperature range. On the other side, a change of the area is not theoretically expected while being in the weak scattering limit\cite{ToyozawaPTP58} and thus we believe the observed decrease of the amplitude of the peak (see Fig.~\ref{SpettriAlfaZoom}), measurable with a better accuracy than the width, to be an indirect evidence of the increase of the width of the line itself.
\subsection{Relation with the Urbach absorption tail}\label{urbachsection}
In the frame of Toyozawa's theory the characteristic Urbach exponential absorption tail (UT) given by Eq.~\ref{urbach} has been predicted to be a typical feature existing on the left wing of Lorentzian exciton peaks as a consequence of the creation of momentary self-trapped excitons.\cite{SumiJPSJ71,SchreiberJPSJ82c} Thus, on a qualitative basis present findings on the delocalized nature of the 10.4~eV exciton agree with the characteristic presence in SiO$_2$ of the UT as observed by absorption measurements between 8~eV and 9~eV.\cite{SaitoPRB2000,VellaPRB09} Furthermore, present conclusions on the negligible influence of amorphous disorder on the 10.4~eV peak are consistent with previous investigations on the effect of structural disorder on the properties of the UT in a-SiO$_2$ showing that, if compared to thermal disorder, structural disorder accounts for a minor portion only of the broadening of the intrinsic edge.\cite{SaitoPRB2000,VellaPRB09}

The agreement becomes much stronger when one notices that the value of the parameter $\hbar\omega_0$ we obtained from the temperature dependence of the 10.4~eV peak is in a very good agreement with that already available in literature\cite{SaitoPRB2000} for a-SiO$_2$, $\hbar\omega_0$=$(0.089\pm0.008)$~eV, deduced from the study of the temperature dependence of the UT.\footnote{This value was obtained from UT data on a-SiO$_2$ samples featuring a low Si-OH impurities content and may be slightly different in other types of a-SiO$_2$. However, the comparison is fully meaningful as the a-SiO$_2$ type used in the present study coincides with that in which $\hbar\omega_0$ was determined in ref. \cite{SaitoPRB2000}} This strongly supports Toyozawa's interpretation of the UT being a low energy feature peculiar of a weakly scattered exciton absorption peak\cite{SumiJPSJ71,SchreiberJPSJ82} and its variations as a function of temperature as a consequence of those of the first exciton peak. In this context it should be noted that, as far as the Urbach tail is concerned, a-SiO$_2$ is not representative of the behavior of all amorphous systems. Indeed, a "glassy" variant of the Urbach tail is observed in several amorphous solids, where the slope of the exponential profile in a semi-logarithmic scale, the so-called Urbach energy, does not depend on temperature in contrast with Eqs.~\ref{urbach} and \ref{sigma}. Thus, the fact that, notwithstanding its amorphous structure, a-SiO$_2$ shows the standard Urbach rule (Eqs.~\ref{urbach} and \ref{sigma}) further indicates that structural disorder hardly affects the electronic properties of the material.

The association of the UT to the 10.4~eV peak in SiO$_2$ has been questioned,\cite{GodmanisPSS1983} based on the position of the temperature-independent crossing point ($E_0$, $\alpha_0$) of the UTs, which is expected to exist as a consequence of Eq.~\ref{urbach}. Indeed, experimental investigations have found the spectral position of the crossing point to be located at $E_0$=9.1~eV in c-SiO$_2$ and at $E_0$=8.7~eV in a-SiO$_2$.\cite{GodmanisPSS1983} Based on the difference between $E_0$ and the position 10.4~eV of the exciton peak it was proposed that the exciton responsible for the 10.4~eV peak is different from that associated to the Urbach tail.\cite{GodmanisPSS1983,TrukhinJNCS92} However, the convergence of the Urbach tails close to the energy position of the main peak is expected only when the energy gap between valence and conduction bands is direct, as in alkali halides.\cite{SumiJPSJ71,Toyozawa} In contrast, if the gap is indirect (i.e. the $K$$=$$0$ exciton level is not at the bottom of the exciton band) the excitonic peak moves to a higher energy, while the crossing point of the UTs is still expected to stay at the energy corresponding to the bottom of the exciton band, i.e., recalling that the width of the exciton energy band is 2$B$ (see section~\ref{theory}), an energy $B$ below the band center.\cite{SumiJPSJ71,Toyozawa} Thus, in the simplest case of a $K$$=$$0$ level located at the center of the band the distance between the converging point and the excitonic peak is $B$. From the experimental point of view this is the situation found for instance in AgCl, where $E_0$=3.3~eV corresponds to the indirect band-gap of the material, while the excitonic peak is at 5.0~eV.\cite{TutihasiPR1957,ToyozawaPTP58,SumiJPSJ71}

Then, for a-SiO$_2$ there is no conflict between $E_0$=8.7~eV and the excitonic peak at 10.4~eV: this disagreement simply leads to conclude that SiO$_2$ features an indirect gap near 9~eV while the direct gap is above the value of $E_P^0$ found here, i.e. close to $\sim$11~eV.\footnote{The separation between the direct gap and $E_P^0$ equals the exciton binding energy.} The indirect nature of the gap in SiO$_2$ is also consistent with results obtained by theoretical calculations.\cite{Pantelides,ChelikowskyPRB77} Based on these considerations, we can roughly estimate $B$ as the separation in energy between $E_0$ and $E_P^0$, i.e. $B$$\sim$10.8-8.7=2.1~eV. While this estimate applies to the simplest case of $K$$=$$0$ level at the center of the exciton band, a lower threshold for $B$ can be obtained if one considers the case of $K$$=$$0$ at the top of the band, and is half of the just calculated value.

\subsection{Parameters affecting exciton dynamics}\label{parameters}
A comparison between present data and the features of the UT in a-SiO$_2$ allowed us to estimate the value $B$$\sim$2.1~eV. Another totally independent line of reasoning leads to a consistent estimate of the magnitude of the same parameter. Several studies on the Urbach tail have led to estimate the parameter (see Eq.~\ref{urbach}) $\sigma_0$$\sim$0.5 for a-SiO$_2$ and $\sim$0.6 for c-SiO$_2$.\cite{Erice} As explained in the introductory section, $\sigma_0$ is basically the inverse of the EPC constant $g$$=$$E_{LR}/B$, leading to $g$$\sim$2 in a-SiO$_2$. Now, the parameter $E_{LR}$, defined in section~\ref{theory} as the relaxation energy within the excited electronic state of the exciton, can be estimated approximately as the half Stokes shift between exciton absorption and emission. Since STE emission is at $\sim$2.5~eV for a-SiO$_2$ and at 2.6~eV for c-SiO$_2$\cite{ItohPRB89,TrukhinJNCS92,Erice} (singlet and triplet emission being separated by a fraction of eV\cite{TrukhinJPCM08}), we get $E_{LR}$$\sim$(10.4-2.5)/2$\sim$4~eV in a-SiO$_2$. Hence, from $g$$=$$E_{LR}/B$ we finally obtain $B$$\sim$2~eV. The good agreement with the previously estimated $B$$\sim$2.1~eV strongly confirms the applicability of the whole line of reasoning. Similarly in c-SiO$_2$ we get $B$$\sim$2.4~eV, $E_{LR}$=4~eV. In the following we will assume $B$$\sim$2~eV both in a-SiO$_2$ and in c-SiO$_2$. The proposed estimates of the parameters $B$, $E_{LR}$ and $g$ are reported in Tab.~\ref{table:res}.

We are now in a position to provide and discuss approximate estimates of all the parameters controlling exciton dynamics. Starting from our estimation of the $B$ values, one can evaluate $D$ from the observed FWHM of the bands. In fact, the first-order theoretical expression FWHM=$\pi$$D$$^2$/$B$ (see section~\ref{theory}) can be inverted to get $D$$\sim$0.74~eV in a-SiO$_2$ (FWHM=0.85~eV) and $D$$\sim$0.66~eV in c-SiO$_2$ (FWHM=0.69~eV). The similarity between these two values confirms that the contribution of amorphous structural disorder to excitonic energy fluctuations is very poor, as anticipated. In the following we will consider $D$$\sim$0.7~eV in both systems, leading to $D/B$$\sim$0.35. The estimated values are reported in Tab.~\ref{table:res}. While, strictly speaking, this value of $D/B$ cannot be considered as $\ll$1, it anyway most assuredly falls into the weak scattering category, also considering the fact that the transition between weak and strong scattering is a gradual one. Fluctuations ($D$) of the exciton energy of the order of 0.7~eV actually imply a rather efficient coupling of excitons with vibrational modes, being the weak scattering condition anyway ensured by the large extension of the excitonic band. This is the situation referred to as strong EPC ($g$$\sim$2) and weak scattering ($D/B$$\sim$0.35), characterized by an excitonic peak which is relatively wide (since FWHM$\propto$$D^2$) yet Lorentzian and by STE-type emission, as expected when $g$$>$1 and observed in SiO$_2$. It should be noted here that another estimate of $D$ can be obtained via Eq.~\ref{allvsT} in the limit $T$$\rightarrow$$0$ by using E$_{LR}$$\sim$4~eV and $\hbar\omega_0$=0.083~eV found before. This yields $D$$\sim$0.6~eV in a-SiO$_2$ at low temperatures. The good agreement with the previously calculated value $D$$\sim$0.7~eV further strengthens our conclusions, especially if one considers that the latter was inferred by an expression for the Lorentzian FWHM which represents only a first-order approximation.

\begin{figure}
\includegraphics[width=8cm]{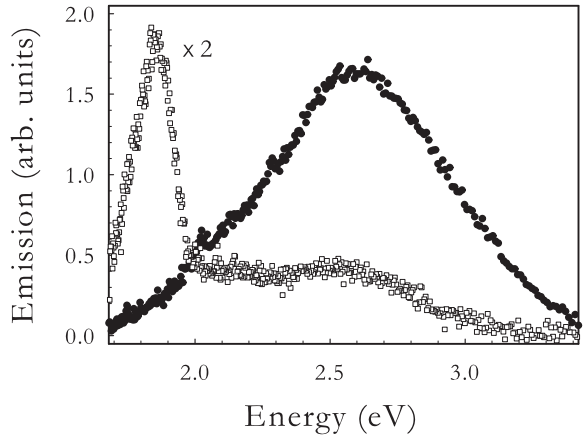}
\caption{\label{fig6rev} Photoluminescence emission spectrum detected at T=10 K in c-SiO$_2$ (circles) and a-SiO$_2$ (squares) under excitation with synchrotron radiation of 10.2 eV photon energy.}
\end{figure}

Being the absorption energy as high as 10.4~eV, a value of $E_{LR}$ of the order of a few eV is reasonable in the scheme of strong EPC. On the other hand, by applying Dexter criterion for luminescence,\cite{ToyozawaPAC1997} the value of $E_{LR}$$\sim$4~eV should be associated to a significant probability of non-radiative excitons decay, typically leading to the creation of point defects.\cite{Nalwa,SkujaSPIE01} This is consistent with the experimentally observed emission properties of SiO$_2$, where non-radiative decay of excitons has been widely recognized as an important mechanism for photo-induced generation of defects.\cite{Nalwa,SkujaSPIE01} In this respect, we report in Fig.~\ref{fig6rev} a typical photoluminescence emission spectrum detected at T=10 K in c-SiO$_2$ and a-SiO$_2$ upon excitation at 10.2~eV, falling within the excitonic peak. The signal observed in c-SiO$_2$ consists in a wide band peaked at 2.60~eV which is the well known emission due to the self trapped exciton (STE). The signal in a-SiO$_2$ is composite: the broad band centered at 2.5~eV is due to STE luminescence, while the 1.9~eV peak is due to the point defect known as non-bridging oxygen hole center,\cite{SkujaSPIE01} generated at low temperature in a-SiO$_2$ by synchrotron radiation.\cite{MessinaPRB10} From data in Fig.~\ref{fig6rev} we were able to estimate the excitonic emission quantum yield $\eta$ by comparing the intensity of the signal with that of a system of known quantum yield (sodium salycilate,\cite{TrukhinSSC03} $\eta$=0.5) measured in the same experimental conditions. In this way, we estimated $\eta$(c-SiO$_2$)$\sim$10$^{-2}$ and $\eta$(a-SiO$_2$)$\sim$10$^{-3}$ under excitation at 10.2~eV. These values are $\ll$1 as expected.

\begin{table}[ht]
\caption{Exciton-related parameters extracted from the temperature dependence of the 10.4~eV excitonic peak in a-SiO$_2$ (upper section). Proposed estimates of the parameters defined in section~\ref{theory} controlling exciton dynamics in a-SiO$_2$ (lower section). Correspondent values for c-SiO$_2$ are reported in the text.}
\centering
\begin{tabular}{c c c c c c}
\hline\hline
& &$\hbar\omega_0$(eV) & $E_P^0$(eV) & $\Delta_0^0$(eV) & \\ [0.5ex]
\hline
& &0.083 & 10.79 & -0.38 &\\
\hline \hline
B(eV) & D(eV) & D/B & E$_{LR}$(eV) & $g$ &$\eta$(10.2~eV) \\ [0.5ex]
\hline
$\sim$2 & $\sim$0.7& $\sim$0.35& $\sim$4 & 2.1&$\sim$10$^{-3}$\\
\hline \hline
\label{table:res}
\end{tabular}
\end{table}

\subsection{Comparison with literature on SiO$_2$ and other systems}\label{sectionD}
The strong experimental evidence that the peak is accurately described by a simple Lorenztian function seriously casts doubt on the model widely accepted in SiO$_2$ literature claiming a strong localization and consequently a low mobility of excitons in a-SiO$_2$.\cite{TrukhinJNCS92,Erice} Such a model, based on the study of the energy transfer processes from excitons to luminescent impurities, described the 10.4~eV band, even in c-SiO$_2$, as the overlap of several Gaussian sub-bands associated to several types of localized excitons. Aside from the papers supporting this model, the other existing literature on SiO$_2$ did not address the issue of the line-shape of the 10.4 eV peak, likely due to insufficient quality of data or to their work being focused on other aspects. In order to further compare present results with previously published ones, we considered data reported by one of the most recent studies of above-edge absorption in SiO$_2$, where measurements were carried out by using a VUV laser plasma light source.\cite{TanPRB05} By carrying out a (not reported) analytical fitting procedure on the absorption spectrum reported for c-SiO$_2$, as acquired from the published paper, we found that the reported 10.5~eV peak in c-SiO$_2$ has a satisfactory Lorentzian profile as well, although the FHWM turns out to be slightly larger (FWHM=0.88~eV) and small deviations appear in the region below 9.3~eV. In regard to a-SiO$_2$ published data do not contrast with ours, although the signal-to-noise ratio is insufficient to conclusively determine the line-shape of the peak.

Lorentzian exciton line-shapes have been reported in literature for several crystalline systems at low temperatures.\cite{TomikiJPSJ73,BurlandJCP77} Although most of them actually feature very sharp absorption and free-exciton photoluminescence emission lines, a few examples exist featuring relatively wide yet Lorentzian exciton line-shapes. These are systems featuring a strong EPC but a weak scattering (as here proposed for SiO$_2$),\cite{SchreiberJPSJ82,ToyozawaPAC1997} and basically consist in wide band-gap crystals such as LiF and NaF. For instance, the first excitonic absorption in LiF and NaF have been reported to be Lorentzians peaked at 12.6~eV and 10.7~eV with 0.4~eV and 0.12~eV widths respectively.\cite{SanoJPhysSocJapan69,PiacentiniSSC1975}$^,$\footnote{The peak in NaF is actually a doublet composed by two Lorentzian shapes. The 0.12 eV FWHM quoted in the text is the width of each of the two components} The analogies between SiO$_2$ and these other very-wide band-gap systems also extend to excitonic photoluminescence properties and to the properties of the far low energy tail of the excitonic peak. In fact, also LiF and NaF feature wide Stokes-shifted self-trapped-exciton (STE) luminescence bands (e.g. emission at 3.4~eV for LiF\cite{NakonechnyiJPCM06}), thus being consistent with the emission properties of SiO$_2$ upon above-edge excitation. Furthermore, also LiF and NaF feature the UT on the left wing of their Lorentzian absorption peaks.\cite{SanoJPhysSocJapan69,TomikiJPhysSocJapan69} Thus, LiF and NaF can be classified as SiO$_2$ as featuring strong coupling to, but weak scattering by phonons.

Excitonic absorption bands in LiF and NaF feature a certain degree of asymmetry which is apparently not observed here for SiO$_2$. Theory predicts asymmetry properties to depend on the position of the $K$$=$$0$ level inside the exciton band.\cite{ToyozawaPTP58} If the $K$$=$$0$ level is at the center of the band, the resulting line-shape is predicted being symmetric, while if the $K$$=$$0$ level is in the upper or lower half of the band, the line-shape is expected being steeper on the right (negative asymmetry) or left (positive asymmetry) side respectively.\cite{ToyozawaPTP58} The last case includes the situation of a direct band-gap (as in LiF and NaF), for which the Lorentzian description is most accurate on the right half of the peak. Actual systems showing Lorentzian profiles with both positive, negative, and very low asymmetry were observed in literature. For instance, AgCl (indirect gap at 3.3~eV) features an excitonic peak at 5.0~eV with 0.3~eV FWHM which is Lorentzian on the left side, while on the right side the overlap with other bands prevent accurate analysis,\cite{TutihasiPR1957,ToyozawaPTP58,SumiJPSJ71} which is a very similar situation as that observed here for SiO$_2$. Thus, in addition to the distance between the crossing point of the UTs and the excitonic peak, also the lack of asymmetry suggests SiO$_2$ to feature an indirect band-gap, and in particular the $K$$=$$0$ level to be close to the center of the exciton band. On the other side, we cannot rule out the possibility of a steeper high-energy wing of the exciton peak, being this hidden by the superposition with the 11.5~eV peak, which would lead to the $K$$=$$0$ level being in the upper half of the excitonic band.

As far as the influence of the amorphous nature of the material is concerned, it is worth noting that the possibility of delocalized excitons was tentatively proposed in other systems, such as a-As$_2$S$_3$,\cite{IgashiPRL81} based on indirect evidences arising from luminescence measurements. On the other side, the behavior of SiO$_2$ is deeply different from that of other materials such as Si, where a dramatic change of the shape and broadening of the excitonic peak at 3.5~eV are observed when going from the crystal to the amorphous solid.\cite{AspnesPRB84} Thus, it cannot be taken for granted for all systems that amorphous disorder introduces only a small perturbation of the delocalized exciton states.
\subsection{Discussion of the higher energy peaks}\label{sectionE}
While there is a general agreement on the fact that the $\sim$14.1~eV and $\sim$17.3~eV peaks are associated to maxima of the inter-band transition cross section,\cite{TanPRB05} as to the $\sim$11.6~eV peak several authors suggested the presence of two or more contributions both peaking in this energy region, only one of which associated to excitonic absorption,\cite{PlatzoderPSS68,ChelikowskyPRB77,BosioEPL92} possibly related to the first excited state of the 10.4~eV exciton.\cite{BosioEPL92} As for the region of the inter-band transition peaks, our data in Fig.~\ref{SpettriAlfa} indicate that the absorption coefficient at $E$$>$12.2~eV slightly increases as a whole without changing its shape from 10 to 150~K. While we acknowledge that present results do not allow to clarify the nature of this increase, it certainly has consequences on the apparent temperature dependence of the $\sim$11.6~eV structure. As a matter of fact, a comparison between Fig.~\ref{SpettriAlfaZoom}-(c) and Fig.~\ref{SpettriAlfa} clearly reveals that the apparent variations of the peak at $\sim$11.6~eV from 150 to 10~K are only a consequence of those observed in the remaining higher energy part of the spectra. On the other hand, in the range of temperatures from 300 to 150~K the observed decrease (Fig.~\ref{SpettriAlfaZoom}-(b)) of the amplitude of the peak at $\sim$11.6~eV is peculiar of this structure only: indeed, it can be clearly singled out from the temperature-induced modifications of the higher energy side of the spectra, since the high and low energy wings of the band do not change with varying temperature from 300 to 150~K. While a quantitative analysis of the temperature dependence of the $\sim$11.6~eV peak is hindered by its overlap with the nearby structures, its decrease for T$>$150~K appears to resemble that observed for the peak at $\sim$10.4~eV in the same range of temperatures: based on these qualitative similarities we argue that in a-SiO$_2$ the peaks at $\sim$10.4~eV and at $\sim$11.6~eV share a common physical origin, signally excitonic.
\section{Conclusions}
The reflectivity spectra of a-SiO$_2$ in the range of temperatures from 10 to 300~K were studied and the correspondent above-edge absorption coefficient was obtained via Kramers-Kronig dispersion analysis. In the absorption spectra four main peaks (at $\sim$10.4~eV, at $\sim$11.6~eV, at $\sim$14.1~eV and at $\sim$17.3~eV) were observed and their temperature dependencies examined. Our results significantly improve the current understanding of excitonic properties of SiO$_2$. We demonstrated that the peak at $\sim$10.4~eV has a remarkably good Lorentzian profile in the whole observed range of temperatures. This result can be interpreted as strong evidence of a weak exciton-phonon scattering and of the excitons in a-SiO$_2$ retaining their mobility properties albeit both the disorder peculiar of its amorphous structure and a strong exciton-phonon coupling. The study of the temperature dependence of the position of the peak at $\sim$10.4~eV allowed to determine the value of the average vibrational energy of phonons interacting with excitons [$\hbar\omega_0$$=$$(0.083\pm0.002)$~eV]. The properties of the Urbach exponential absorption tail and those of self-trapped exciton photoluminescence excited above-gap are qualitatively and quantitatively consistent with those of the 10.4~eV absorption peak. This allows to interpret self-consistently the whole set of optical properties related to excitation of a-SiO$_2$ from $\sim$8~eV to $\sim$11~eV as determined by excitons, demonstrates the applicability of Toyozawa's theory for excitonic line-shapes to an amorphous system and supports the interpretation of the Urbach tail as resulting from momentary self-trapping of excitons. Moreover, our data provide estimates of the main theoretical parameters governing exciton-phonon interaction. Exciton dynamics turns out to be satisfactorily described in the context of a strong exciton-phonon coupling coexisting with weak exciton-phonon scattering, with SiO$_2$ featuring an indirect band-gap at $\sim$9~eV and a threshold for direct band-to-band transitions at $\sim$11~eV. a-SiO$_2$ and c-SiO$_2$ are found to be very similar in regard to exciton properties, apart from a small additional contribution to the excitonic energy fluctuations in the amorphous case due to disorder. Finally, the qualitative similarities between the temperature dependencies of the peaks at $\sim$10.4 and $\sim$11.6~eV in a-SiO$_2$ suggest they both have an excitonic origin.
\section*{Acknowledgements}
We acknowledge financial support received from DESY and we thank the LAMP group (http://www.fisica.unipa.it/amorphous) for useful and stimulating discussions.

\section*{References}
\providecommand{\newblock}{}


\begin{thebibliography}{10}
\expandafter\ifx\csname url\endcsname\relax
  \def\url#1{{\tt #1}}\fi
\expandafter\ifx\csname urlprefix\endcsname\relax\def\urlprefix{URL }\fi
\providecommand{\eprint}[2][]{\url{#2}}

\bibitem{Nalwa}
Nalwa H~S (ed) 2001 {\em Silicon-based Materials and Devices\/} (Academic
  Press, USA) ISBN 0-12-513909-8

\bibitem{SkujaSPIE01}
Skuja L, Hosono H and Hirano M 2001 {\em Proc. SPIE\/} {\bf 4347} 155

\bibitem{SkujaSPIE03}
Skuja L, Hosono H, Hirano M and Kajihara K 2003 {\em Proc. SPIE\/} {\bf 5122} 2

\bibitem{KajiharaJCSJ07}
Kajihara K 2007 {\em J. Ceramic Soc. Jap.\/} {\bf 115} 85

\bibitem{ChangPRL2000}
Chang E~K, Rohlfing M and Louie S~G 2000  {\bf 85} 2613

\bibitem{VellaPRB08}
Vella E, Boscaino R and Navarra G 2008  {\bf 77} 165203

\bibitem{PhilippSSC66}
Philipp H~R 1966 {\em Sol. State Commun.\/} {\bf 4} 73

\bibitem{PhilippJPCS71}
Philipp H~R 1971 {\em J. Phys. Chem. Solids\/} {\bf 32} 1935

\bibitem{LamyAO77}
Lamy P~L 1977 {\em Appl. Opt.\/} {\bf 16} 2212

\bibitem{BosioEPL93}
Bosio C and Czaja W 1993 {\em Europhys. Lett.\/} {\bf 24} 197

\bibitem{TanJACS03}
Tan G~L, Lemon M~F and French R~H 2003 {\em J. Am. Ceram. Soc.\/} {\bf 86} 1885

\bibitem{TanPRB05}
Tan G~L, Lemon M~F, Jones D~J and French R~H 2005  {\bf 72} 205117

\bibitem{WeinbergPRB79}
Weinberg Z~A, Rubloff G~W and Bassous E 1979  {\bf 19} 3107

\bibitem{DiStefanoSSC71}
DiStefano T~H and Eastman D~E 1971 {\em Solid State Commun.\/} {\bf 9} 2259

\bibitem{PlatzoderPSS68}
Platz\"{o}der K 1968 {\em Phys. Stat. Sol.\/} {\bf 29} K63

\bibitem{LaughlinPRB80}
Laughlin R~B 1980  {\bf 22} 3021

\bibitem{ItohPRB89}
Itoh C, Tanimura K, Itoh N and Itoh M 1989  {\bf 39} 11183

\bibitem{BosioEPL92}
Bosio C, Czaja W and Mertins H~C 1992 {\em Europhys. Lett.\/} {\bf 18} 319

\bibitem{AlexandrovNIMA89}
Alexandrov Y~M, Vishnjakov V~M, Makhov V~N, Sidorin K~K, Thrukhin A~N and
  Yakimenko M~N 1989 {\em Nucl. Instr. and Meth. in Phys. Res. A\/} {\bf 282}
  580

\bibitem{EvrardPRB82}
Evrard R and Trukhin A~N 1982  {\bf 25} 4102

\bibitem{TrukhinJNCS92}
Trukhin A~N 1992 {\em J. Non-Cryst. Solids\/} {\bf 149} 32

\bibitem{Erice}
Trukhin A~N 2000 {\em Defects in SiO$_2$ and Related Dielectrics: Science and
  Technology\/} ed Pacchioni G, Skuja L and Griscom D~L (Kluwer Academic
  Publishers, USA) p 235 ISBN 0-7923-6685-9

\bibitem{ChelikowskyPRB77}
Chelikowski J~R and Schl\"{u}ter M 1977 {\em Phys. Rev. B\/} {\bf 15} 4020

\bibitem{Pantelides}
{N F Mott} {1978} {\em {The Physics of SiO$_2$ and its Interfaces}\/} ed {S T
  Pantelides} ({Pergamon, New York}) p~{80}

\bibitem{LohSSC64}
Loh E 1964 {\em Sol. State Commun.\/} {\bf 2} 269

\bibitem{SorokinOS76}
Sorokin O~M and Blank V~A 1976 {\em Opt. Spectrosc.\/} {\bf 41} 353

\bibitem{ToyozawaPTP58}
Toyozawa Y 1958 {\em Progr. Theor. Phys.\/} {\bf 20} 53

\bibitem{SchreiberJPSJ82}
Schreiber M and Toyozawa Y 1982 {\em J. Phys. Soc. Japan\/} {\bf 51} 1528

\bibitem{SumiJPSJ71}
Sumi H and Toyozawa Y 1971 {\em J. Phys. Soc. Japan\/} {\bf 31} 342

\bibitem{SchreiberJPSJ82c}
Schreiber M and Toyozawa Y 1982 {\em J. Phys. Soc. Japan\/} {\bf 51} 1544

\bibitem{MessinaPRL2010}
Messina F, Vella E, Cannas M and Boscaino R 2010  {\bf 105} 116401

\bibitem{Toyozawa}
Toyozawa Y 2003 {\em Optical Processes in Solids\/} (Cambridge University
  Press, UK) ISBN 0-521-55605-8

\bibitem{ToyozawaPAC1997}
Toyozawa Y 1997 {\em Pure \& Appl. Chem.\/} {\bf 69} 1171

\bibitem{Tauc}
Tauc J 1974 {\em Amorphous and liquid semiconductors\/} ed Tauc J (Plenum
  Press, London and New York) p 159

\bibitem{Ashcroft}
Ashcroft N~W and Mermin N~D 1981 {\em Solid State Physics\/} (Holt-Saunders,
  International Editions) ISBN 0-03-049346-3

\bibitem{LevitonSPIE2006}
Leviton D and Frey B 2006 {\em Proc. of SPIE\/} {\bf 6273} 62732K

\bibitem{MatsuokaJNCS1991}
Matsuoka J, Kitamura N, Fujinaga S and Kitaoka T 1991 {\em J. Non-Cryst.
  Solids\/} {\bf 135} 86

\bibitem{BosioCM1989}
Bosio C, Harbeke G, Czaja W and Meier E 1989 {\em Condens. Matter\/} {\bf 62}
  748

\bibitem{KuboJMathPhys63}
Kubo R 1963 {\em J. Math. Phys.\/} {\bf 4} 174

\bibitem{SaitoPRB2000}
Saito K and Ikushima A~J 2000  {\bf 62} 8584

\bibitem{VellaPRB09}
Vella E and Boscaino R 2009  {\bf 79} 085204

\bibitem{GodmanisPSS1983}
Godmanis I, Trukhin A~N and H\"{u}bner K 1983 {\em Phys. Stat. Sol.\/} {\bf
  116} 279

\bibitem{TutihasiPR1957}
Tutihasi S 1957  {\bf 105} 882

\bibitem{TrukhinJPCM08}
Trukhin A~N 2008 {\em J. Phys.: Condens. Matter\/} {\bf 20} 125217

\bibitem{MessinaPRB10}
Messina F, Vaccaro L and Cannas M 2010  {\bf 81} 035212

\bibitem{TrukhinSSC03}
Trukhin A~N, Kink M, Maksimov Y and Kink R 2003 {\em Solid State Commun.\/}
  {\bf 127} 655

\bibitem{TomikiJPSJ73}
Tomiki T, Miyata T and Tsukamoto H 1973 {\em J. Phys. Soc. Japan\/} {\bf 35}
  495

\bibitem{BurlandJCP77}
Burland D~M, Konzelmann U and Macfarlane R~M 1977 {\em J. Chem. Phys.\/} {\bf
  67} 1926

\bibitem{SanoJPhysSocJapan69}
Sano R 1969 {\em J. Phys. Soc. Japan\/} {\bf 27} 695

\bibitem{PiacentiniSSC1975}
Piacentini M 1975 {\em Solid State Commun.\/} {\bf 17} 697

\bibitem{NakonechnyiJPCM06}
Nakonechnyi S, K\"{a}rner T, Lushchik A, Lushchik C, Babin V, Feldbach E,
  Kudryavtseva I, Liblik P, Pung L and Vasil'chenko E 2006 {\em J. Phys.:
  Condens. Matter\/} {\bf 18} 379

\bibitem{TomikiJPhysSocJapan69}
Tomiki T and Miyata T 1969 {\em J. Phys. Soc. Japan\/} {\bf 27} 658

\bibitem{IgashiPRL81}
Igashi G~S and Kastner M 1981  {\bf 47} 124

\bibitem{AspnesPRB84}
Aspnes D, Studna A and Kinsbron E 1984  {\bf 29} 768

\end{thebibliography}
\end{document}